\documentclass[ floatfix,reprint,
superscriptaddress,
 amsmath,amssymb,
 aps,
prb, longbibliography
]{revtex4-1}

\usepackage{graphicx}
\usepackage[danish,english]{babel}
\usepackage[utf8]{inputenc} 
\usepackage[T1]{fontenc}
\usepackage{amsmath}
\usepackage{color}
\newcommand{\qqq}{\mathbf{q}}
\newcommand{\QQQ}{\mathbf{Q}}

\newcommand{\GGG}{\mathbf{G}}

\newcommand{\ddd}{\mathbf{d}}
\newcommand{\eee}{\mathbf{e}}

\newcommand{\STO}{SrTiO$_3$}
\begin{document}

\title{Phonon dispersion of quantum paraelectric \STO{} in electric fields}
\author{Henrik Jacobsen}
\email{henrik.jacobsen.fys@gmail.com}
\affiliation{DMSC, Asmussens Allé 305, 2800 Kongens Lyngby, Denmark
}
\affiliation{Nanoscience Center, Niels Bohr Institute, University of Copenhagen, 2100 Copenhagen, Denmark}
\affiliation{Paul Scherrer Institute, Laboratory for Neutron Scattering and Imaging, 5232 Villigen, Switzerland}

\author{Marek Barthkowiak}
\affiliation{Paul Scherrer Institute, Laboratory for Neutron Scattering and Imaging, 5232 Villigen, Switzerland}

\author{Tobias Weber}
\affiliation{Institute Laue-Langevin, 71 Avenue des Martyrs, CS 20156, 38042 Grenoble cedex 9, France}

\author{Uwe Stuhr}
\affiliation{Paul Scherrer Institute, Laboratory for Neutron Scattering and Imaging, 5232 Villigen, Switzerland}
\author{Bertrand Roessli}
\affiliation{Paul Scherrer Institute, Laboratory for Neutron Scattering and Imaging, 5232 Villigen, Switzerland}
\author{Christof Niedermayer}
\affiliation{Paul Scherrer Institute, Laboratory for Neutron Scattering and Imaging, 5232 Villigen, Switzerland}
\author{Urs Staub}
\email{urs.staub@psi.ch}
\affiliation{Paul Scherrer Institute, Laboratory for Condensed Matter, 5232 Villigen, Switzerland}

\begin{abstract}
Here we report on an elastic and inelastic neutron scattering study addressing the effect of electric fields on quantum paraelectric SrTiO$_3$. Our elastic scattering results find small changes as a function of field in a superlattice reflection that sample the octahedral rotations, which is indicative of only weak coupling of octahedral rotation and electric polarization. {\color{black} Using inelastic neutron scattering, we measured the dispersion of the transverse acoustic and transverse optic phonon in \STO{} as function of applied electric fields. }By collecting not only the change in gap, but also of the dispersion, we can better quantify the changes in the lattice dynamics. The findings are put in context to recent field DFT calculations predicting the E-field effect on the atomic motions of the lowest lying transverse optical (soft) mode. We find hints of non-linear coupling to the acoustic mode and to the phonon with polarization perpendicular to the E-field, which shows the non-linearity in the chemical potential that is also relevant when strongly drive the SrTiO$_3$ with strong E-field (THz) pulses. 

\end{abstract}

\maketitle
\section{Introduction}
\STO{} is an incipient ferroelectric material that is in the vicinity of a paraelectric-to-ferroelectric phase transition \cite{Muller1979}. The ferroelectric phase is suppressed by quantum fluctuations, and so \STO{} is classified as a quantum paraelectric material\cite{Zhong1996,Shin2021}. Because of the vicinity of this phase transition, the properties of \STO{} are strongly sensitive to external electric fields \cite{Worlock1967,Muller1979,Braeter1989,Sidoruk2016,Li2019}.

In particular, the transverse optical (TO) phonon mode of \STO{} that softens strongly with decreasing temperature, hardens significantly at the zone center upon the application of a moderate electric field, as was demonstrated  using Raman scattering {\color{black}\cite{Worlock1967,Akimov2000}}. {\color{black} The mode is only Raman active in applied electric fields \cite{Fleury1968}.}
In addition, upon cooling below 105 K, the oxygen octahedra in \STO{} tilt, and it undergoes an antiferrodistortive transition from a cubic Pm$\bar{3}$m phase to a tetragonal I4/mcm phase {\color{black} \cite{Shirane1969, Hirota1995}}. The electric field-induced phonon hardening is present below 80 K and is strongest at low temperatures.{\color{black} \cite{Worlock1967,Fleury1968}}

Raman scattering only probes the phonons very close to the zone center. With inelastic neutron scattering, the full phonon dispersion is available and has been investigated previously in great detail for \STO{} in the absence of electric fields\cite{Stirling1972, Yamada1969, Choudhury2008,He2020}. However, the effect of electric fields on the dispersion of the phonons, particularly on the TO soft mode, has not yet been explored.

Much theoretical effort has been dedicated to explaining the field-induced hardening of the TO phonon {\color{black}\cite{Worlock1967,Muller1979,Braeter1989,Torres2019}}. Density functional theory {\color{black}(DFT)} calculations in electric fields are notoriously difficult, but have been employed to describe the hardening of the TO phonon at the zone center \cite{Naumov2005}. {\color{black} We note that another DFT calculation at much higher fields (Ref. \cite{Torres2019}) could not reproduce the hardening of the TO phonon at the $\Gamma$ point}. However, even from a theoretical viewpoint, the effect of an external electric field on the dispersion of the TO mode remains unexplored. 

Recent work driving the soft mode with strong single cycle THz pulses resulted in strong non-linear phonon-phonon couplings \cite{Kozina2019}. Strong anharmonicity has also led to the disappearance of acoustic phonon branches in inelastic neutron scattering experiments\cite{He2020}. 
More recently, a clear hardening of the TO mode was found when the mode was directly excited with THz electric fields, similar as for the application of static electric fields.  \cite{Li2019} In addition, a long-lived polar state appears when driving modes at higher frequencies \cite{Nova2019}. To improve our understanding of phonon hardening in either non-linear excitations or static electric fields, a determination of the field dependent dispersion of the TO mode might lead to additional benchmarks for describing the suppression of quantum fluctuations.   
 
In this paper, we address the anharmonicity in the quantum fluctuation regime by a direct determination of the TO phonon soft mode dispersion using neutron scattering in applied electric fields. We show that the TO phonon is well described by a simple model of optic phonons. Furthermore, we show that the intensity of the transverse acoustic mode increases with field, while that the TO phonon decreases. This indicates that the two modes are coupled. 

\section{Experimental}

We carried out inelastic neutron scattering experiments at the thermal triple-axis spectrometer EIGER \cite{Stuhr2017} and the cold triple-axis spectrometer TASP, both located at the Paul Scherrer Institute (PSI) in Switzerland.\cite{Boni1996}

The sample was a high quality commercial substrate of size $20\times20\times2.2$ mm, with a mass of 4.2 g, obtained from Mateck. 

Electrical contacts were made on the flat surfaces of the sample using silver paste, and connected to a voltage source using a specialized sample stick developed at PSI \cite{Bartkowiak2014}. Voltages of up to 2 kV were applied, corresponding to electric fields of up to about 9 kV/cm. In both experiments, the electric field was applied along the cubic [001] axis, with the sample being inserted in a standard orange cryostat, achieving temperatures down to 2 K. 

At EIGER, the sample was aligned with the cubic [110] and [001] axes in the horizontal scattering plane. The fixed final energy was 14.7 meV,  and we used a PG filter to suppress higher harmonic contamination.

At TASP, the sample was rotated by 45 degrees so that the cubic [100] and [001] axes were in the horizontal scattering plane. Here, the outgoing energy was fixed at 8 meV, and a pyrolytic graphite (PG) filter was placed after the sample. No collimation was used in both experiments.
Wave vectors, $\QQQ$, will be expressed in the cubic setting in units of $(2\pi/a,2\pi/b,2\pi/c)$ with the average low-temperature lattice parameter  $a=b=c=3.89$ \AA{}.

We measured the $q=0$ TO phonon energy using constant $\QQQ$ scans at the (220), (002) and (200) reciprocal space points. The dispersion was measured with constant energy scans along $(H,H,2)$, $(H,0,2)$ and $(2,0,H)$. All inelastic neutron scattering experiments were performed at 2K. 
As the instrumental resolution significantly affects the observed signal, the model described below has been convoluted with the instrument resolution, calculated using Takin 2.3 {\color{black} \cite{Weber2021,Takin}}
with the Popovici algorithm \cite{Popovici1975}.

\section{Results}

\subsection{Octahedral tilts}
We first show the impact of the electric field on the oxygen octahedral tilts upon cooling below the antiferrodistortive transition at 105 K, to test the correlation between octahedral rotation and induced polarization. 

In the tetragonal phase, the (103) nuclear Bragg peak has non-zero intensity. This position corresponds to $\left(\frac{1}{2},\frac{1}{2},\frac{3}{2} \right)$ in cubic notation, and in the cubic phase this peak is forbidden. There is a single additional free fractional oxygen coordinate in the low temperature structure, which is directly related to the rotation of the oxygen octahedra, and therefore the intensity of the (103) reflection is  to first order proportional to the angle of rotation of the  octahedra.  This rotation is monitored by measuring the intensity of the cubic  $\left(\frac{1}{2},\frac{1}{2},\frac{3}{2} \right)$ peak.{\color{black} \cite{Muller1991,Hirota1995}}

\begin{figure}
    \centering
    \includegraphics[width=0.4\textwidth]{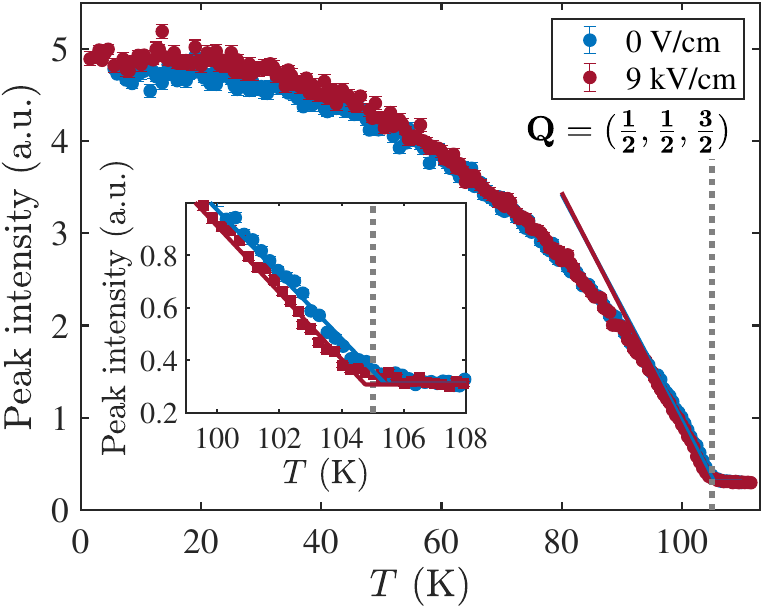}
    \caption{Neutron scattering intensity of the cubic $\left(\frac{1}{2},\frac{1}{2},\frac{3}{2} \right)$ peak as function of temperature in zero and 9 kV/cm electric fields. The peak intensity is a direct measure of the oxygen octahedra tilt angle that occurs in the tetragonal phase, as described in the text.{\color{black} Measurements were obtained down to 5 K in zero field, and 1.5 K in 9 kV/cm applied field.}}
    \label{fig:Fig1}
\end{figure}

In Fig.~\ref{fig:Fig1} we show the neutron scattering intensity of this peak as a function of temperature for both zero field and in 9 kV/cm, while cooling. There is a small difference between the two curves, indicating that the electric field has some impact on the oxygen octahedra tilts. We find $T_c= 105.4(1)$ K and $104.7(1)$ K for 0 V and 9 kV/cm, respectively. This shows that there might be a small coupling between the ferroelectric distortion caused by the electric field and the octahedral rotation as predicted by e.g. the domains walls creating a polarization, which is observed below approximately 80K. \cite{Salje2013a}. The effect at 105 K is rather small, because the dielectric constant is small compared to low temperatures, where a much larger effect of the electric field on the structure occurs. It also shows that the population of orthogonal domains in the tetragonal phase is not affected by the field when cooling through the second-order antiferrodistortive transition, even though theory predicts a clear coupling between ferroelectricity and octhahedral rotations. \cite{Aschauer2014}. The coupling found in theory \cite{Aschauer2014} is obtained indirectly through the change of lattice constants, which affect both the rotation and the polarization.

{\color{black} A neutron diffraction experiment in applied electric fields might shed more light on this coupling.}

\subsection{TO phonon at the zone center}
\begin{figure}
\includegraphics[width=0.37\textwidth]{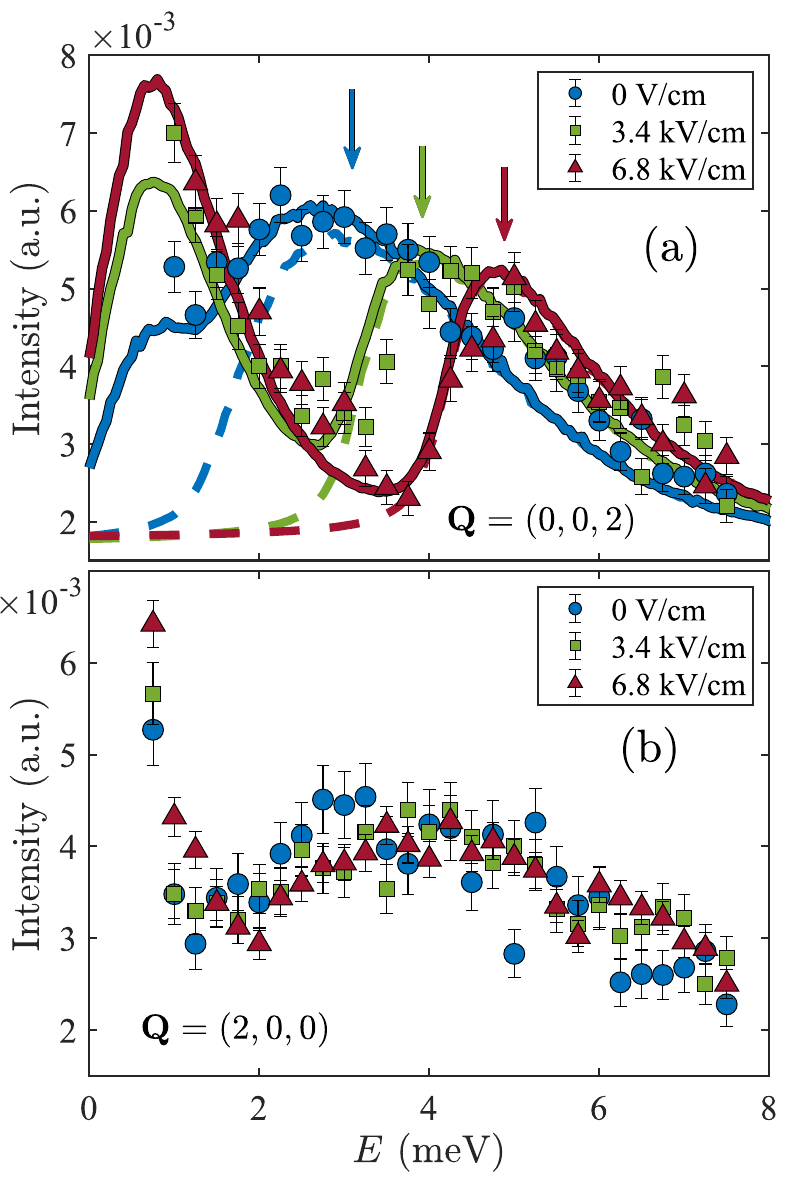}
	\caption{
Neutron scattering intensity as function of energy transfer, $E$, in various electric fields applied along (001), with two directions of the scattering vector.  
(a) At $\QQQ=(002)$, the TO phonon mode is observed as a peak that moves to higher energies with increasing fields, indicated with arrows. At lower energies, the acoustic phonon is seen as a peak. The lines are {\color{black} global fits to the entire dispersion including the data shown in Fig.~\ref{fig:Fig3}} as described in the text.
(b) At $\QQQ=(200)$, the probed TO phonon polarization is perpendicular to the field, and the hardening is much weaker. }
	\label{fig:Fig2}
\end{figure}

We now turn to the TO phonon, starting at the zone center. 
Fig.~\ref{fig:Fig2} shows inelastic neutron scattering data at the $\Gamma$ points $\QQQ=(002)$ and (200) in zero and applied electric fields of 3.4 and 6.8 kV/cm along the (001) direction.

At $\QQQ=(002)$, Fig.~\ref{fig:Fig2}(a), the scattering vector is parallel to the electric field. The TO phonon mode with polarisation parallel to the applied field is seen as a peak in intensity (indicated by arrows), and the hardening with applied field is seen by the peak moving to larger energies, in agreement with Raman data \cite{Worlock1967}. 

In contrast, at $\QQQ=(200)$, Fig.~\ref{fig:Fig2}(b), the scattering vector is perpendicular to the applied field, and the effect of the electric field is much weaker. 

As inelastic neutron scattering is only sensitive to nuclear motion parallel to $\QQQ$, these results indicate that the electric field couples much more strongly to the motion parallel to the field (as expected), leaving the phonons with polarization perpendicular to the field almost invariant. 

We note that the transverse acoustic (TA) phonon is seen as a peak {\color{black} below $\sim1$} meV. In zero field, the signal from the TA phonon overlaps with that of the TO phonon, while at higher fields, the two phonons are well separated. The signal from the TA phonon also overlaps with the tails of the Bragg peak. The intensity of the TA phonon increases with increasing field, which we shall discuss later. 

\subsection{TO phonon dispersion}
We next study  the dispersion of the TO phonon in electric field. To do this, we used TASP to measure the TO phonon along $\QQQ=(H,0,2)$ at multiple constant energies at various electric fields. Examples of our data are presented in Fig.~\ref{fig:Fig8}. Two phonon branches are observed, seen as peaks in the spectra that disperse in opposite directions from the zone center as the energy increases. At energies below 5 meV, the signal from the two TO {\color{black} peaks} overlap, and the resolution of the instrument must be included in the fitting process. We therefore fit all the data simultaneously to an overall model of the dispersion, with electric field-dependent parameters describing the dispersion. We outline the model below.

As shown in Fig.~\ref{fig:Fig2}, both the TA and TO phonons contribute to the measured signal. Our model therefore includes both the TO and TA phonon modes. 

\begin{figure}
\includegraphics[width=0.4\textwidth]{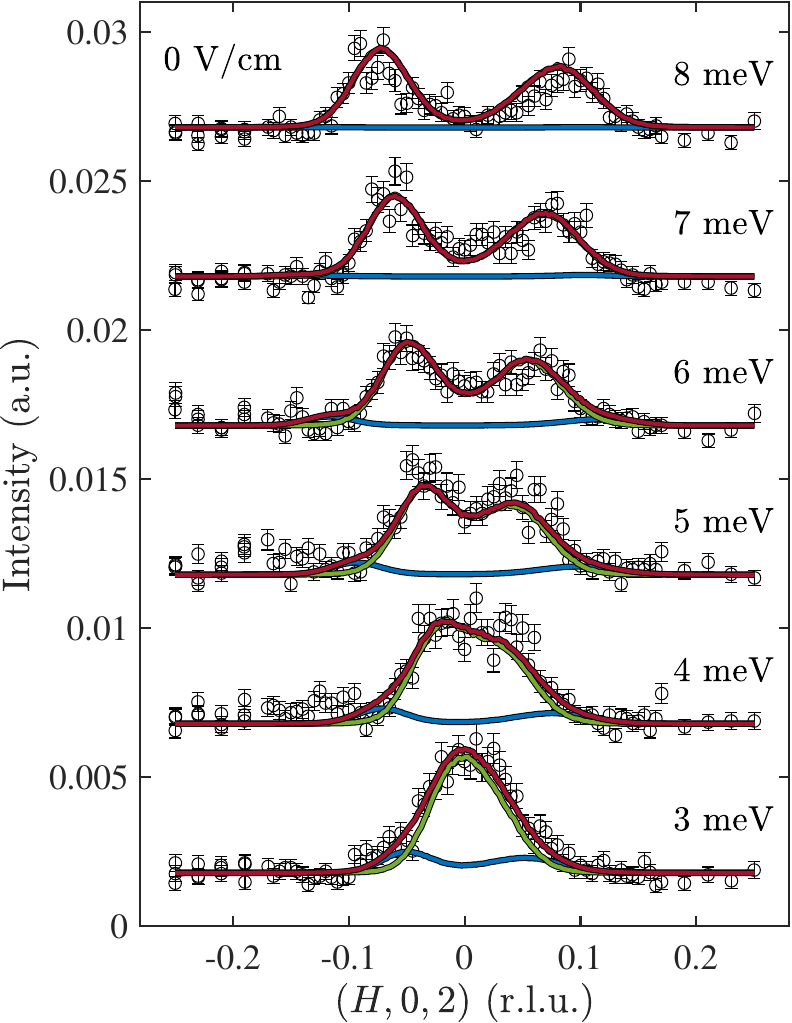}
\caption{Neutron scattering intensity along ($H,0,2$) for various energies in zero field, measured at TASP. The TO phonon mode is clearly visible as two peaks that disperse as the energy is increased. We fit all measurements simultaneously to a single global model, as described in the text. 
The solid red lines show the total fit, while the blue and green lines show the contributions from the TA and TO phonons, respectively. }
\label{fig:Fig3}
\end{figure}

We model the neutron scattering cross-section of the phonons as \cite{Squires2012}

\begin{align}
\left(\frac{d^2 \sigma}{d\Omega dE}\right)^\text{TA/TO}=&S^\text{TA/TO}(\QQQ) \sum_s |\QQQ \cdot \eee_{s}|^2 \left(\frac{n(\omega_s)+1}{ \omega_s} \right)\times \nonumber \\ 
&\delta(\QQQ-\qqq-\GGG) \frac{\gamma/\pi}{\gamma^2+(\omega-\omega_s)^2},
\end{align}
where TA/TO labels the type of phonon, $n(\omega)$ is the Bose occupation factor, 
$\QQQ$ is the scattering vector, $\qqq$ is the phonon propagation vector and $\GGG$ is a reciprocal lattice vector. $s$ runs over the branches of the phonon modes, and $\eee$ is the polarization of the phonon mode. We assume the line shape of both phonon excitations is a Lorentzian in energy, with half width at half maximum $\gamma$, corresponding to a finite lifetime of the phonon, $\tau\sim \hbar/\gamma$. Finally, $S^\text{TA}(\QQQ)=S_0^\text{TA}$ is a constant, which includes the structure factor of the Bragg peak associated with the phonon. From our fits described below, we find in the region of interest that $S^\text{TO}(\QQQ)\approx \omega_s S_0^\text{TO}$.
See the appendix for details about the model. 

From the modelling shown in Fig.~\ref{fig:Fig3}, it is evident that the acoustic phonon contributes significantly up to about 5 meV.

The dispersion of the TO phonon is well approximated by a slight modification of the dispersion of a diatomic chain, phase shifted to place the minimum at $\Gamma$:
\begin{align}
    \hbar \omega^\text{TO} \approx \Delta/2+\sqrt{(\Delta/2)^2+C^2\sin^2(D|\qqq|)},
\end{align}
where $\Delta$ is the energy gap at $\qqq=0$.

At each electric field, we simultaneously fitted the constant $\QQQ$ scan and the measurements at all energy transfer to the model described above, convoluted with the instrumental resolution using Takin. To account for background, we added  an additional term $K_1+K_2E$. The free parameters were the overall intensity scale of the TA and TO phonons, $S_0^A$ and $S_0^O$, the dispersion parameters $\Delta$ and $C$, and the background parameters. The phonon life-times, seen as the (inverse) energy width of the phonons, have very little influence on the fits, and so we fixed them at $\gamma=0.2$ meV as found in previous neutron scattering experiments \cite{Yamada1969,Bruce1983}.

For the TA phonon, we found from preliminary analysis that only the intensity scale varied significantly with increasing electric field. At low fields, the signal from the TA and TO phonons overlaps at low energies, and the two contributions can be difficult to disentangle. On the other hand, at high fields, the hardening of the TO phonon means that most of the low-energy scattering comes from the TA phonon, as seen e.g. in Fig.~\ref{fig:Fig2}(a). We therefore first fitted the high field measurements, allowing also the parameters describing the dispersion of TA phonon to vary. To improve the robustness of the fit at lower fields, we subsequently fixed these two parameters. 

Our fit in zero field is shown in Figs.~\ref{fig:Fig2} and \ref{fig:Fig3}, and we find excellent agreement with all our data. 

In applied electric fields, the {\color{black} doubly} degenerate TO phonon is lifted due to the preferred orientation given by the direction of the field. To describe the data in applied fields, we fixed the background parameters and parameters for the modes with polarization perpendicular to the electric field to the values found in zero field, and allowed only the optic and acoustic modes parallel to the electric field to vary. We again find excellent agreement with all our data. These fits are shown in the appendix, Fig.~\ref{fig:Fig8}.

\begin{figure}
\includegraphics[width=0.45\textwidth]{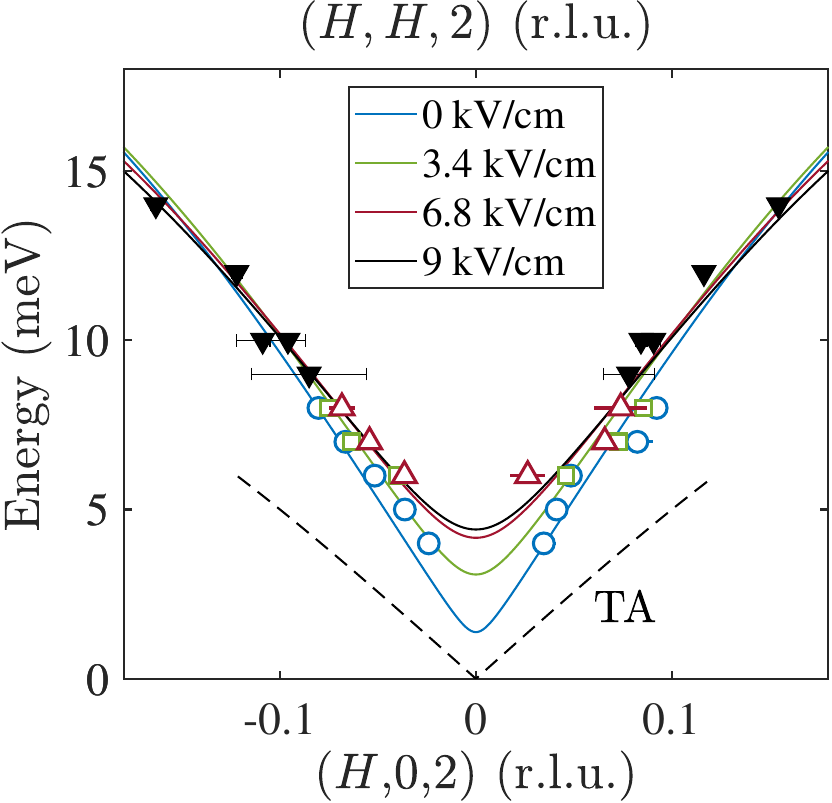}
\caption{The dispersion of the TO phonons with polarization parallel to field taken at 2 K for selected electric fields.  As the fits are global fits to the spectra, we note that the points represent approximate peak positions found by fitting the data to two Gaussians. {\color{black} The 9 kV/cm data (black triangles) were measured along $(H,H,2)$ as indicated by the upper x-axis. The dashed line shows the fitted dispersion of the TA phonon.}}
\label{fig:Fig4}
\end{figure}

From the fits, we obtain the dispersion of the TO mode, with findings being summarized in Fig.~\ref{fig:Fig4}. In zero field, the dispersion is quite steep in the vicinity of the  $\Gamma$ point, reminiscent of an acoustic mode. This can be seen as a precursor of the ferroelectric phase transition, at which this polar soft mode should go to zero. As the electric field is increased, the phonon becomes less steep, with a reduced slope, representing a reduced group velocity of the mode. This is a natural effect of the increase of the gap for a soft mode. However, it is not necessarily generally true, e.g. in the presence of a Kohn anomaly.
.

The data at the highest applied electric field, 9 kV/cm, were obtained at EIGER, with the sample rotated to measure along $(H,H,2)$. The fits to our data from EIGER are shown in Fig. \ref{fig:Fig7} of the appendix, and once again, the agreement with our model is excellent.

\begin{figure}
\includegraphics[width=0.35\textwidth]{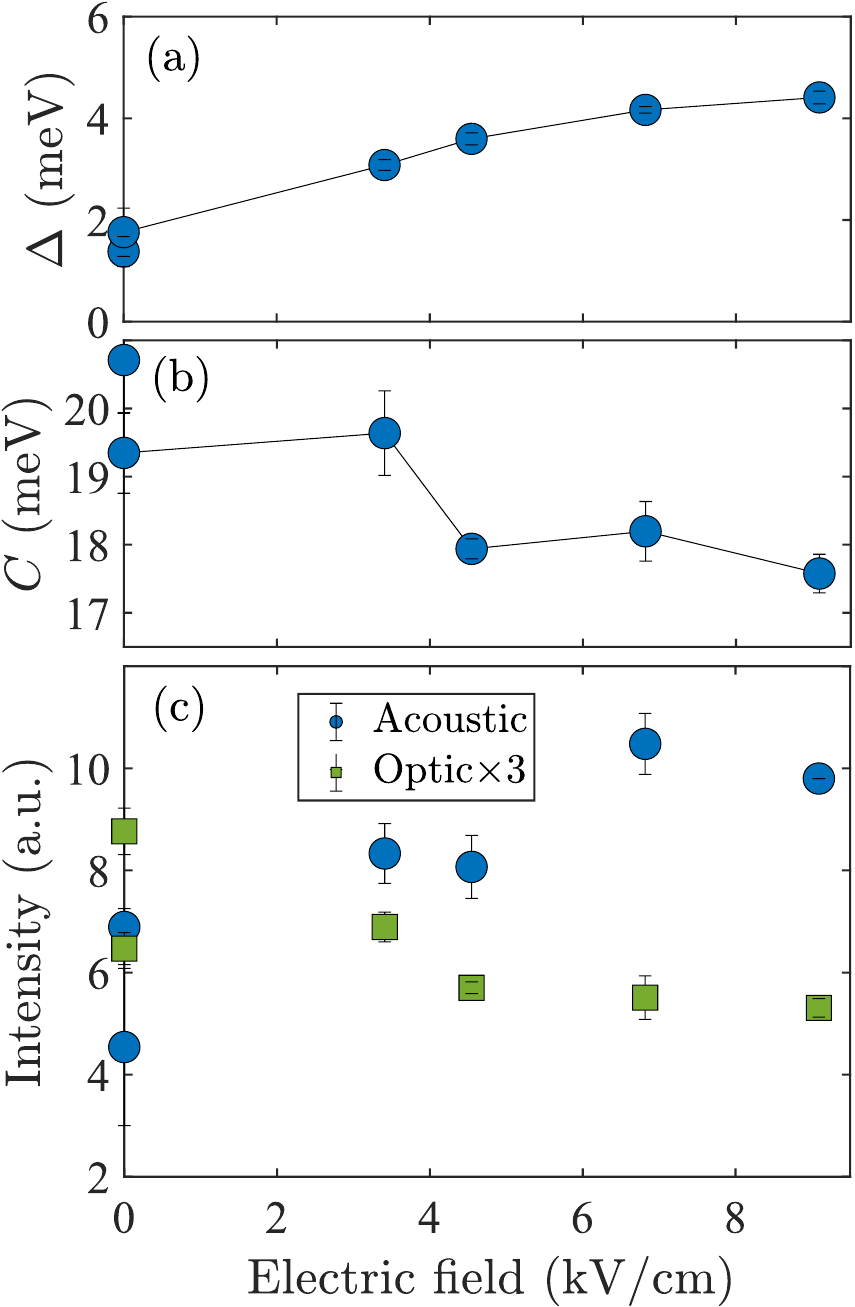}
\caption{
(a,b) The electric field dependence of the parameters $\Delta$, and $C$.
(c) The intensities of the acoustic and optic phonons as a function of electric field.}
\label{fig:Fig5}
\end{figure}

The main result of this analysis is the electric field dependence of the parameters describing the phonon dispersion. These are summarized in Fig.~\ref{fig:Fig5}. We see that $\Delta$ increases smoothly with field, which at $\QQQ=0$  corresponds to the gap at the $\Gamma$ point increasing as expected.  $C$ decreases slightly, which would correspond to a slight change of the sound velocity if $\Delta$ would be zero. 

As the electric field is increased, the intensity of the TO phonon decreases, while that of the TA phonon increases. 
This shift of spectral weight between the two modes indicates that they are coupled, as indeed found by Fauque et al. \cite{Fauque2022a} in a recent zero-field study. That study also found an anomalous dispersion of the TA mode at very low energies. Unfortunately, we cannot resolve this small effect in our study. 

The hardening of the TO phonon is strongest at the zone center, and nearly non-existent at the zone boundaries. 
First-principles density functional theory has been successfully used to describe how the phonon hardens with applied electric fields \cite{Naumov2005}, although only the hardening at the zone center was calculated and the used field strengths were much larger than in our experiment. 

{\color{black}
The applied electric field predominantly hardens the phonon with polarization parallel to the field. This is because the electric field creates changes to the atomic position along the field direction. In the harmonic approximation  (without mode coupling), only the force constant in that direction should change. However, the calculations by Ref.~\cite{Naumov2005} indicate that the orthogonal phonon should harden at sufficiently large electric fields of about $10^5$V/cm.}

In our TASP experiment, where we measured the phonon perpendicular to the applied electric field, we could only reach about 6.8 kV/cm, Fig.~\ref{fig:Fig2}, about an order of magnitude less. The phonon hardening is indeed much weaker for the phonon orthogonal to the field, but there is a significant difference between the data at 6.8 kV/cm compared to the zero field data, with the latter peaking at lower energies. Our measurements therefore confirm the theory by Naumov et al.,~\cite{Naumov2005} that the electric field hardens the TO mode perpendicular to the applied field. We also note that the orthogonal acoustic mode is influenced by the field, as clearly seen in Fig.~\ref{fig:Fig2}(b), where the scattering intensity around 1 meV energy transfer is significantly larger at 6.8 kV/cm compared to zero field. It would be interesting to follow this effect to even larger electric fields.

In a larger context, we can also compare the energy shift of the TO mode and the induced polarization from the electric field with experimental observations from the application of ultra short electric field pulses. Such experiments, using single cycle THz pulses on SrTiO$_3$, have recently been interpreted in terms of induced ferroelectricity occurring in the quantum paraelectric regime at low temperatures. \cite{Li2019} This manifested in the observation of second harmonic generation (SHG) signals with “long” decay times. Using molecular dynamic simulations, polarization of 0.06 to 0.08 C/m$^2$ have been predicted caused by the E-field pulses of several 100 kV/cm amplitude. The creation of a polar distortion, either by an effective field (as seen in our data) or by a “quasistatic” order, would lead to a stiffening of the soft mode. From the above predicted size of the polarization, one can estimate the required effective E-field from the dielectric constant \cite{Muller1979} to be in the order of 4 kV/cm.  This size of E-field leads to a shift of the soft mode to roughly doubling its frequency (fig \ref{fig:Fig4}). Indeed, for lower fields, a frequency shift has been observed in the THz experiments, \cite{Li2019} however, with a much more complicated substructure for higher fields.

Further THz pump studies have been performed on the structurally similar but more simple KTaO$_3$ system \cite{Cheng2023}. They presented 
an alternative interpretation of the enhanced SHG signal when exciting with THz pulses. The study claimed that the SHG signal originates from the polarization of randomly oriented polar nano domains. The clearly observed stiffening of the soft mode has been associated to the non-linear potential. The advantage of this interpretation is, either the effective field from the pump or the induced polarization should create a time dependence in the soft mode frequency, which would be interesting to be tested in further time resolved experiments. It is respect, it remains interesting to further understand the correlation of the effects of static applied E-fields with that of non-linear behaviour of ultra short dynamic E-fields.

\section{Conclusion}

We have measured the dispersion of SrTiO$_3$ in several applied electric fields. We find that the hardening of the TO phonon is strongest at the zone center and weak at the edges. We have confirmed that the hardening is mainly in the phonon with polarization parallel to the electric field, but also that the effect perpendicular to the field is non-zero, in agreement with theory.
The TO and TA phonons are coupled, and spectral weight is transferred to the acoustic phonon with increasing electric fields. 

 It also shows that the population of orthogonal domains in the tetragonal phase is not affected by the field when cooling through the antiferrodistortive transition, even though theory predicts a clear coupling between ferroelectricity and octhahedral rotations. \cite{Aschauer2014}

\begin{acknowledgments}
We thank N. Spaldin and T. Esswein for illuminating discussions. HJ was funded by the  EU  Horizon  2020  programme under the Marie Sklodowska-Curie grant agreement No701647, and by the Carlsberg Foundation. This work is based on experiments performed at the Swiss spallation neutron source SINQ, Paul Scherrer Institute, Villigen, Switzerland.

\end{acknowledgments}

\newpage
\appendix
\section{Phonon model}

We here briefly outline the phonon model. The neutron scattering cross-section of a phonon on the neutron energy loss side is {\color{black} \cite{Boothroyd2020}}

\begin{align}
\frac{d^2 \sigma}{d\Omega dE}=\frac{k_f}{k_i} \frac{(2\pi)^3}{v_0} \sum_{\GGG,s} |G_s(\QQQ)|^2 \times \\
\left(\frac{n(\omega_s)+1}{2 \omega_s} \right) \delta(\QQQ-\qqq-\GGG)\delta(\omega-\omega_s),
\end{align}
where $\QQQ$ is the scattering vector, $\qqq$ is the phonon propagation vector and $\GGG$ is a reciprocal lattice vector. $s$ runs over the phonon modes, $k_f$ and $k_i$ are the final and initial neutron wave vector, respectively, $v_0$ is the volume of the unit cell, and $G_s$ is the structure factor of the phonon:
\begin{align}
    G_s(\QQQ) = \sum_d \frac{\bar{b}_d}{\sqrt{m_d}} \QQQ\cdot \eee_{ds} \exp(-W_d) \exp(i \QQQ\cdot \ddd).
\end{align}
Here, $d$ runs over the atoms in the unit cell, $b$ is the scattering length, $m$ the mass, $\eee$ is the polarization of the phonon mode and $W$ the Debye-Waller factor. {\color{black} We note that the standard notation used here can be confusing, since $\GGG$ refers to a reciprocal lattice vector, while $G_s$ refers to the phonon structure factor.}

We need the phonon structure factor mainly to interpret the different modes we observe.  The $k_f/k_i$ factor is cancelled by the monitor efficiency scaling with $k_i$. 

\section{More data}
We here show all of our raw data along with our model as explained in the main text.

\begin{figure}[b]
\includegraphics[width=0.48\textwidth]{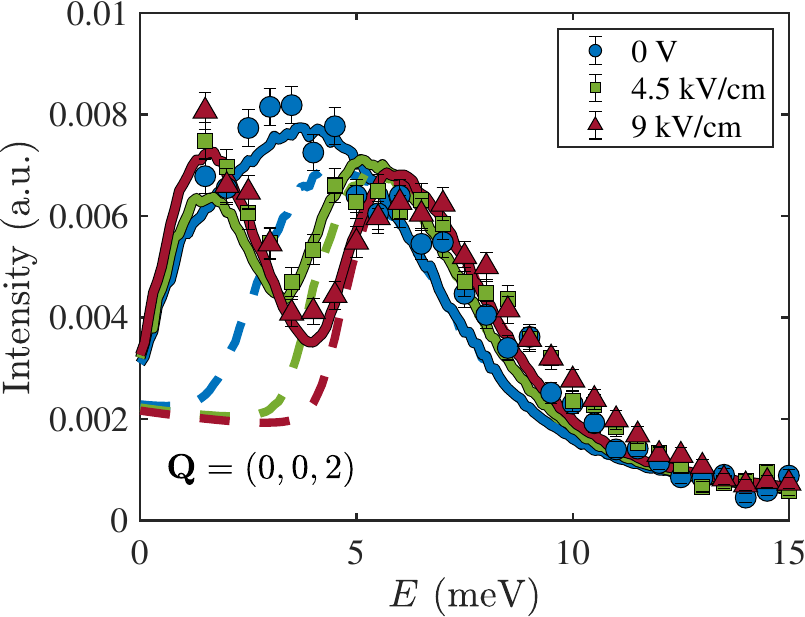}
\caption{
Neutron scattering intensity as function of energy transfer, $E$, in various electric fields applied along (001), measured at ${\bf Q}=(002)$. The TO phonon mode is observed as a peak that moves to higher energies with increasing fields, indicated with arrows. At lower energies, the acoustic phonon is seen as a peak. The lines are fits as described in the text.
 }
\label{fig:Fig6}
\end{figure}

Fig.~\ref{fig:Fig6} shows our constant ${\bf Q}=(002)$ measurement at EIGER at 0 V/cm, 4.5 kV/cm and 9 kV/cm.

\begin{figure*}[]
\includegraphics[width=0.4\textwidth]{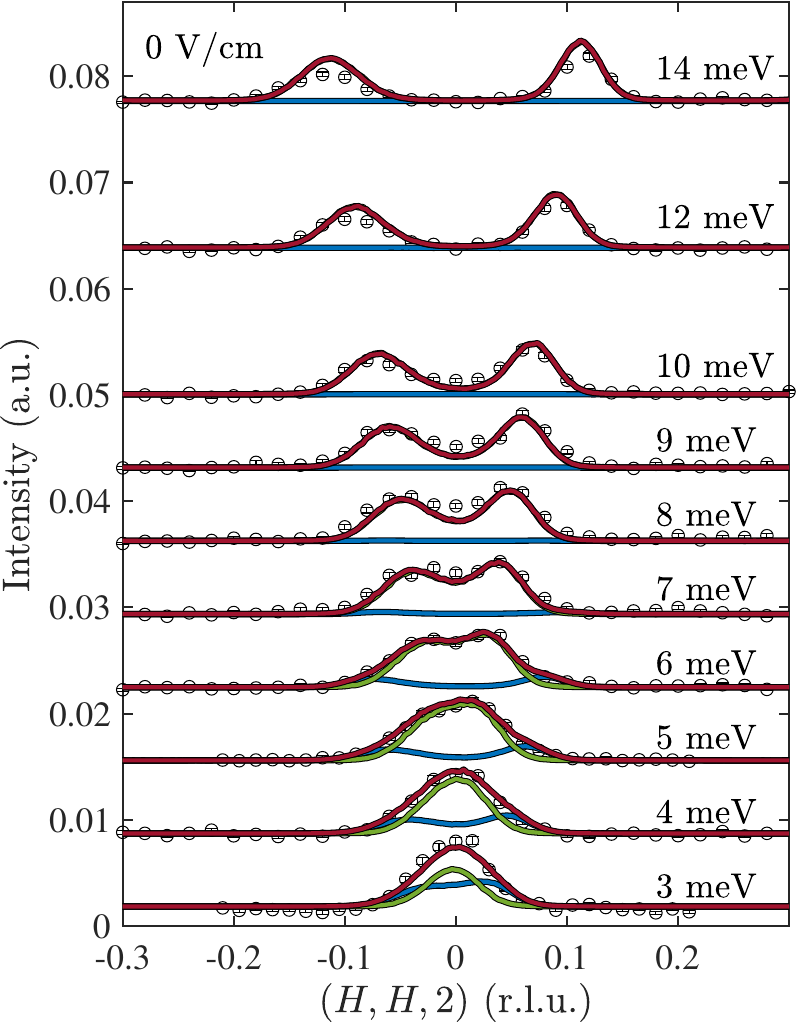}
\includegraphics[width=0.4\textwidth]{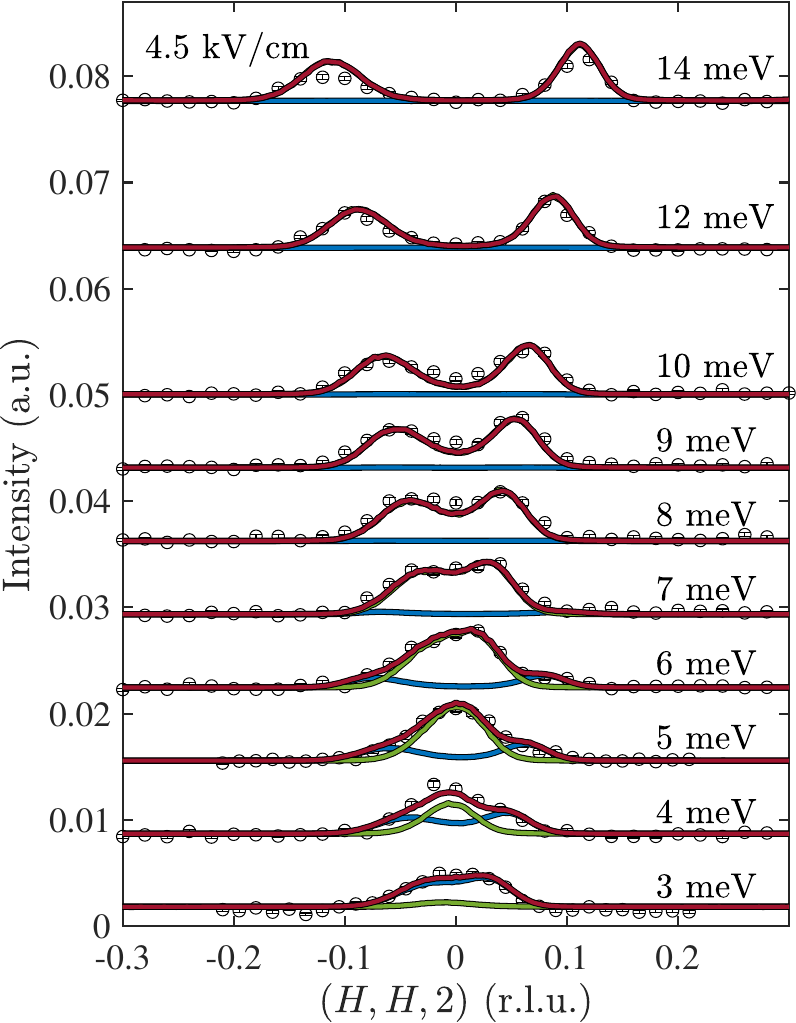}
\includegraphics[width=0.4\textwidth]{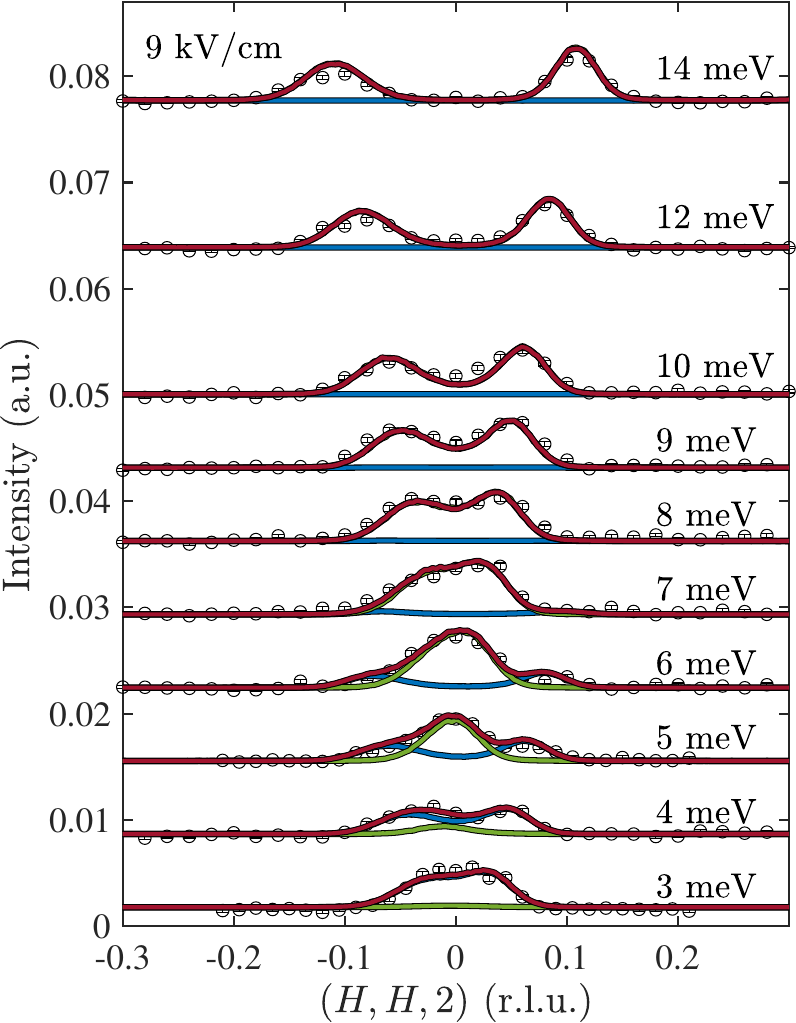}
\caption{Neutron scattering intensity along ($H,H,2$) for various energies in (top left) 0 kV/cm, (top right) 4.5 kV/cm and (bottom) 9 kV/cm. The TO phonon mode is clearly visible as two peaks that disperse as the energy is increased. We fit all measurements simultaneously to a single global model, as described in the text. 
The solid red lines show the total fit, while the blue and green lines show the contributions from the TA and TO phonons, respectively. At large fields, only the TA phonon contributes to the low energy scattering.}
\label{fig:Fig7}
\end{figure*}

Fig.~\ref{fig:Fig7} shows our constant energy measurements from EIGER data at 0 V/cm, 4.5 kV/cm and 9 kV/cm. 

\begin{figure*}[ht]
\includegraphics[width=0.4\textwidth]{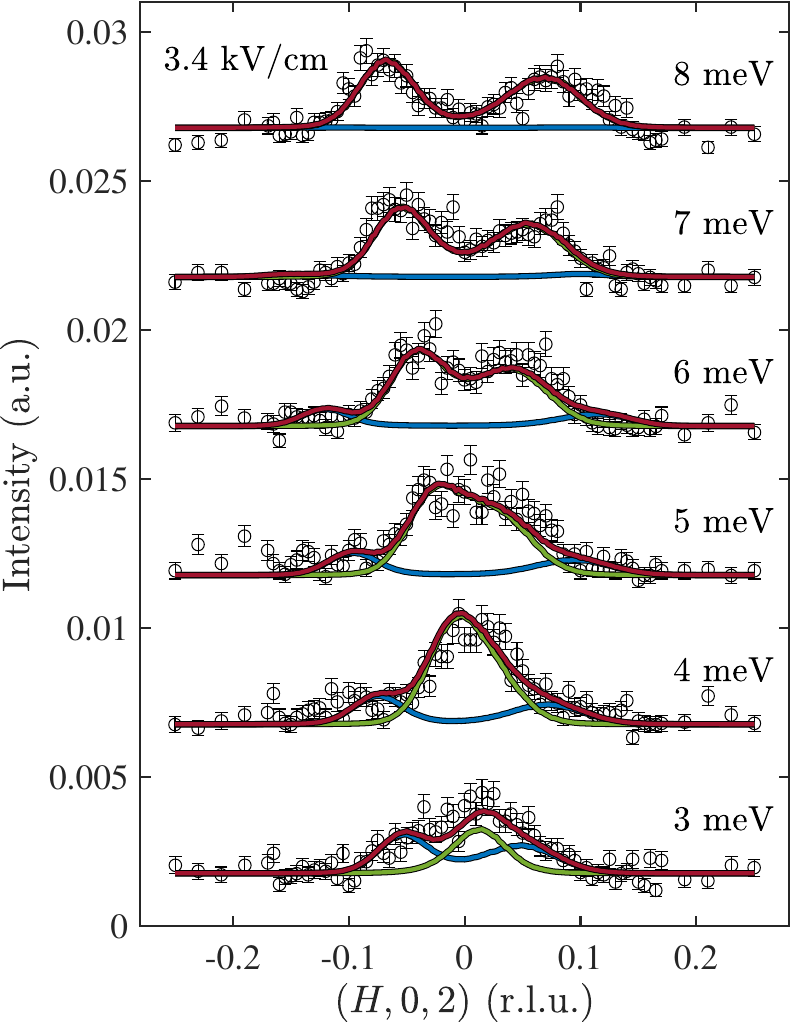}
\includegraphics[width=0.4\textwidth]{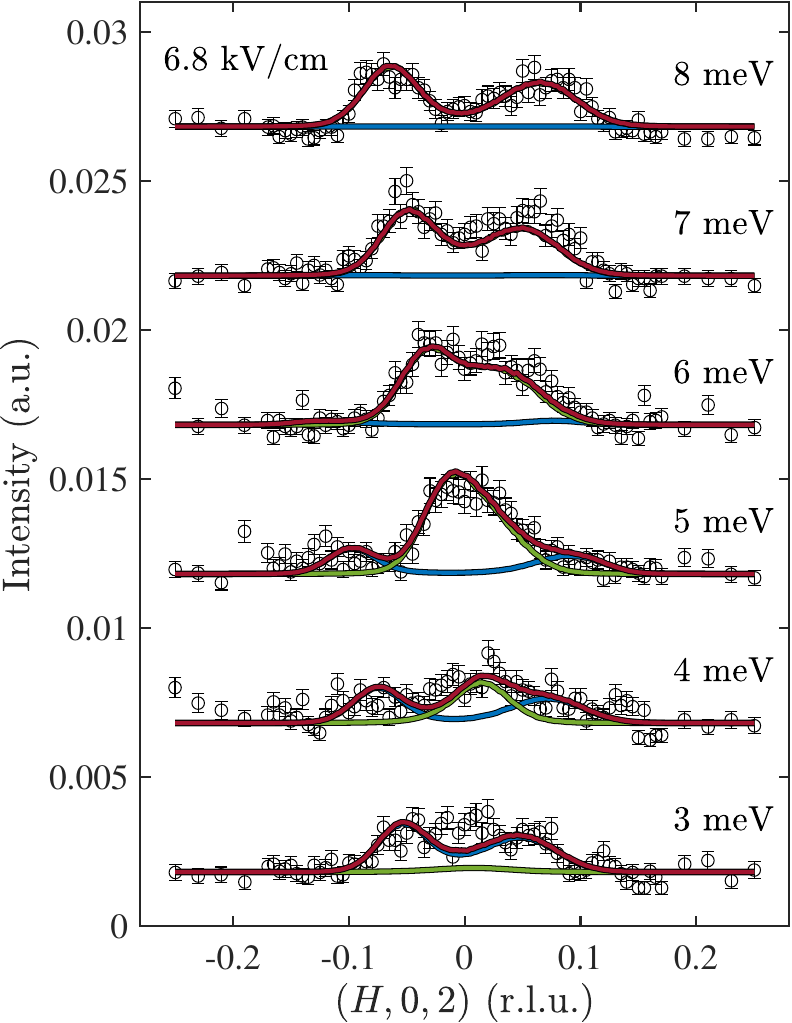}
\caption{Neutron scattering intensity along ($H,0,2$) for various energies in (left) 3.4 kV/cm and (right) 6.8 kV/cm. The TO phonon mode is clearly visible as two peaks that disperse as the energy is increased. We fit all measurements simultaneously to a single global model, as described in the text. 
The solid red lines show the total fit, while the blue and green lines show the contributions from the TA and TO phonons, respectively. At large fields, only the TA phonon contributes to the low energy scattering.}
\label{fig:Fig8}
\end{figure*}

Fig.~\ref{fig:Fig8} shows our constant energy measurements from TASP at 3.4 kV/cm and 6.8 kV/cm. 

\clearpage 

\end{document}